\documentstyle[aps,preprint]{revtex}
\input epsf
\tightenlines
\begin{document}
\draft
\preprint{DUKE-TH-96-126}
\title{Chemical Equilibration of an Expanding Quark-Gluon Plasma}
\author{Dinesh Kumar Srivastava, Munshi Golam Mustafa}
\address{Variable Energy Cyclotron Centre, 1/AF Bidhan Nagar, Calcutta 
700 064}
\author{Berndt M\"{u}ller}
\address{Department of Physics, Duke University, Durham, North Carolina
27708-0305}
\date{\today}
\maketitle

\begin{abstract}
The chemical equilibration of the parton distribution in collisions of 
two heavy nuclei at the CERN Large Hadron Collider is investigated.
Initial conditions are obtained from a self-screened 
parton cascade calculation. The onset of transverse expansion of
the system is found to impede the chemical equilibration. 
The system initially approaches chemical equilibrium, but then is 
driven away from it, when the transverse velocity becomes large. 
\end{abstract}
\narrowtext

It is expected that very highly energetic heavy ion collisions lead 
to the formation of a new phase of strongly interacting matter,
called the quark-gluon plasma (QGP).  At collider energies it seems 
reasonable to visualize the two nuclei as two clouds of valence and 
sea partons which pass through each other and interact \cite{klaus}. 
The multiple parton collisions are thought  to produce a dense plasma
of quarks and gluons.  After its formation this plasma will expand, 
cool, and become more dilute.  If quantum chromodynamics admits a 
first-order deconfinement or chiral symmetry phase transition, it is 
likely that the system will pass through a mixed phase of quarks, 
gluons, and hadrons, before the hadrons lose thermal contact and stream 
freely towards the detectors.  

Several questions about the structure of the matter formed in these
nuclear collisions arise.  Does the initial partonic system attain 
kinetic equilibrium?  Probably yes, as the initial parton density is 
large, forcing the partons to suffer many collisions in a very short 
time \cite{kin}.  Does it attain chemical equilibration?  This will 
depend on the time available \cite{biro,3a,3b} to the partonic system 
before it converts into a mixed phase.
The time available for this is perhaps  too short (3--5 fm/c) at the
energies ($\sqrt{s} \leq 100$ GeV/nucleon) to be reached at the 
Relativistic Heavy Ion Collider (RHIC). At the energies ($\sqrt{s} 
\leq 3$ TeV/nucleon) that will be achieved at the CERN Large Hadron
Collider (LHC) this time could be large (more than 10 fm/$c$).
If one considered only a longitudinal expansion of the system, 
the QGP formed at LHC energies would approach chemical equilibrium 
very closely, due to the higher initial temperature predicted to be 
attained there.  However, the life-time of the plasma would also be 
large enough to allow a rarefaction wave from the surface of the plasma 
to approach the center.  We shall see that this could have interesting 
consequences for the evolution of the plasma.

It has recently been shown \cite{sspc} that color screening provides
a mechanism for the elimination of infrared divergences in the
partonic cascades following the interaction of two heavy nuclei.
Early, hard scatterings produce a medium which screens the longer
ranged color fields associated with softer interactions. When two
heavy nuclei collide at high enough energy, the screening occurs
on a length scale where perturbative QCD still applies.
This approach yields predictions for the initial conditions of the
forming QGP without the need for cut-off parameters. The resulting 
values to be adopted in the present study are listed in Table I.
They show that the QGP initially is extremely hot, with temperatures
around 1 GeV, but not chemically equilibrated.

Once the kinetic equilibrium has been attained we  assume that
the system can be described by the equation for conservation of 
energy-momentum of an ideal fluid:
\begin{equation}
\partial_\mu T^{\mu \nu}=0 \; , \qquad
 T^{\mu \nu}=(\epsilon+P) u^\mu u^\nu + P g^{\mu \nu}
\label{hydro}
\end{equation}
where $\epsilon$ is the energy density and $P$ is the pressure measured 
in the frame comoving with the fluid. We use natural units ($\hbar = c =1$).
The four-velocity vector $u^\mu$ for the 
fluid satisfies the constraint $u^2=-1$. For a partially equilibrated 
plasma of massless particles the equation of state can be written as
\begin{equation}
\epsilon=3P=\left[a_2 \lambda_g +  b_2 \left (\lambda_q+\lambda_{\bar q}
\right ) \right] T^4
\label{eos}
\end{equation}
where $a_2=8\pi^2/15$, $b_2=7\pi^2 N_f/40$, $N_f \approx 2.5$ is 
the number of dynamical quark flavors, and $\lambda_k$ is the fugacity
for the parton species $k$.  We take the speed of sound $c_s=1/\sqrt{3}$
as implied by (\ref{eos}) and solve the hydrodynamic equations 
(\ref{hydro}) for a system undergoing a boost invariant longitudinal 
($z$-axis) and cylindrically symmetric transverse expansion (see Ref.
\cite{vesa} for details) with initial conditions obtained earlier. It 
is sufficient to solve the problem for $z=0$, because of the assumption of 
boost invariance.  Confining ourselves \cite{biro} to the dominant 
reactions $gg \leftrightarrow ggg$, and $gg \leftrightarrow q\bar{q}$  
for the equilibration of the partons, we write the  master equations 
governing the parton densities as \cite{biro}
\begin{eqnarray}
\partial_\mu (n_g u^\mu)&=&n_g(R_{2 \rightarrow 3} -R_{3 \rightarrow 2})
                    - (n_g R_{g \rightarrow q}
                       -n_q R_{q \rightarrow g} ) \nonumber\\
\partial_\mu (n_q u^\mu)&=&\partial_\mu (n_{\bar{q}} u^\mu)
                     = n_g R_{g \rightarrow q}
                       -n_q R_{q \rightarrow g},
\label{master1}
\end{eqnarray}
in an obvious notation.  These equations lead to  partial differential
equations for the fugacities:
\begin{eqnarray}
\frac{\gamma}{\lambda_g}\partial_t \lambda_g &+& \frac{\gamma v_r }{\lambda_g}
\partial_r \lambda_g +\frac{1}{T^3}\partial_t (\gamma T^3) + \frac{v_r}{T^3}
\partial_r (\gamma T^3)  \nonumber\\ 
&+& \gamma \partial_r v_r +\gamma \left( \frac{v_r}{r}+\frac{1}{t}\right) 
\nonumber\\ &=& 
R_3 ( 1- \lambda_g ) -2 R_2 \left( 1-\frac{\lambda_q \lambda_{\bar{q}}}
{\lambda_g^2}\right) \, , 
\nonumber\\
\frac{\gamma}{\lambda_q}\partial_t \lambda_q &+& \frac{\gamma v_r }{\lambda_q}
\partial_r \lambda_q +\frac{1}{T^3}\partial_t (\gamma T^3) + \frac{v_r}{T^3}
\partial_r (\gamma T^3) \nonumber\\
&+& \gamma \partial_r v_r +\gamma \left( \frac{v_r}{r}+\frac{1}{t}\right) 
\nonumber\\ &=&
R_2 \frac{a_1}{b_1} \left(
\frac{\lambda_g}{\lambda_q}-\frac{\lambda_{\bar{q}}}{\lambda_g}\right)\, ,
\label{master}
\end{eqnarray}
where $R_2$ and $R_3$ are related to the rates appearing in (\ref{master1})
and are given by, 
\begin{eqnarray}
R_2 & \approx & 0.24 N_f \alpha_s^2 \lambda_g T \ln (1.65/\alpha_s \lambda_g)
\nonumber\\
R_3 & = & 1.2 \alpha_s^2 T (2\lambda_g-\lambda_g^2)^{1/2},
\end{eqnarray}
$v_r$ is the transverse velocity and $\gamma=1/\sqrt{1-v_r^2}$.  These will
reduce to (34, 35) of Bir\'o et al.~\cite{biro}, if there is no transverse 
expansion of the system.  The hydrodynamic equation (\ref{hydro}) is 
numerically solved to get $\epsilon(r,t)$ and $v_r(r,t)$ which are  then 
used to solve the equations for the fugacities \cite{vesa}. We have 
verified that our results near $r=0$ closely follow the results for a 
purely longitudinal expansion, till the time when the rarefaction wave 
has not reached the center. We have solved the hydrodynamic equations 
by assuming that the initial transverse velocities are zero. 

Recall that for a boost invariant longitudinal expansion (\ref{hydro})
provides that $\epsilon \, \tau^{4/3}$ is a constant\cite{bj}.
Thus in order to bring out
the consequences of transverse expansion in a transparent manner, we
plot the constant energy density contours $\epsilon(r,t)=\epsilon_i/N^{4/3}$
for N=1,6,11..,71 (Figure 1a), which would be lines parallel to the N=1 case 
extending upto $r=R_T$  and separated by $\Delta N \,t_i$, if there were no
transverse expansion of the system.  In the present case N=71 corresponds to 
an energy density\cite{com} of 1.45 GeV/fm$^3$ below which we assume
the description in terms of a perturbative QCD plasma to fail.  We have 
also shown a line $r=R_T-c_s t$ which indicates the radial distance of 
the region affected by the rarefaction wave at any given time $t$.
We see that almost the entire fluid is likely to be affected by the flow
as the life time of the system is sufficiently large.  A more rapid cooling 
of the system is also indicated by the decreasing separation of the contours 
at later times at, say, $r=0$. This will affect the evolution of the 
chemical equilibration in several competing ways.

Thus in Figs. 1b and 1c we show the fugacities for the gluons and 
the quarks, respectively, along some of the above constant energy density 
contours.  Several interesting observations can be made. We note that the 
quark fugacities lag behind the gluon fugacities at all times and all 
radial distances. This is not surprising, considering the lower starting 
values for $\lambda_q$.  Recalling that increasing $N$ denotes passage of 
time, we see that the fugacities at moderate values of $r$ first
increase with passage of time. However beyond $N=31$, ($t \approx$ 8 
fm/$c$, near $r=0$) when the transverse expansion of the system is large, 
it starts decreasing again, except at $r=0$ where it continues to increase, 
as all the radial derivatives in (\ref{master}) are negligible there. 

This can be understood by noting that the evolution of the partonic
density which is governed by $\partial_\mu (n u^\mu)$ consists of
two terms:  $u^\mu \partial_\mu n$ which is the rate of change of 
the density in the comoving frame, and $n\, \partial_\mu u^\mu$ 
which is the rate of change of the density due to the expansion of 
the fluid element in the comoving frame \cite{str}.  Once the 
transverse expansion of the fluid  starts developing, the second term 
and also the radial derivatives in (\ref{master})
grow very rapidly and drive the system away from chemical equilibrium. 

The fugacities decrease monotonically with increase in $r$ which is not
surprising because the transverse flow effects grow with increasing
$r$ and also the time available to the system in the QGP phase shrinks
at larger $r$.  This can have  very interesting consequences, the most 
intriguing of which is the possibility of the formation of a hot and rare 
partonic system near the surface as compared to a cooler but denser 
partonic system at the center.  Thus, e.g., we estimate that the 
temperature at the end of our QGP phase is about 200 MeV at $r$=6 fm, 
but only about 160 MeV at $r$=0 fm. This variation in the temperature 
gets more severe at RHIC energies \cite{future}, as the life-time of the
QGP phase is much smaller there, thus terminating the journey of the 
partonic system far away from chemical equilibrium. One may create 
arbitrary competitions between the life-time and the transverse flow 
effects by considering systems of varying dimensions and initial 
conditions.

It is not clear whether such a system will go through a mixed phase, 
say by nucleation of a hadronic droplet \cite{joe}, at all radial 
distances even if QCD admits a first order quark-hadron phase transition. 
Lowering the energy density at which we terminate our calculations will 
only sharpen this difference as the transverse flow will further enhance 
the gradients.

The development of the transverse flow and its
consequences on the chemical equilibration are sensitively dependent
on the evolution of the transverse velocity of the fluid $v_r(r,t)$.
In Fig.~2 we have shown the variation of the transverse velocity along
some of the constant energy density contours shown in Fig.~1a. We see that
the transverse velocity develops rapidly and spreads to smaller radii
very quickly, as the speed of sound which drives the pressure is
assumed to be rather large. It is conceivable that the speed of sound
is reduced when the energy density gets smaller\cite{boyd}. The life-time of
the system would, however, be larger then, and thus the system remains prone
to the consequences of the transverse expansion, as seen earlier. 

If we now assume that the matter goes to a mixed phase (at small radii, 
at least), the transverse velocity attained at the end of the QGP phase 
will stay constant for a comparatively long time, and it may be possible 
to devise means to actually determine it from the measured particle spectra.  
Since D-mesons and even B-mesons are expected to be copiously produced at
LHC energies, the decay leptons from these heavy mesons could serve as
indicators of the collective transverse flow, because the
leptons will inherit a substantial fraction of the significant kinetic
energy carried by their parents. Due to their large mass and relatively
small cross sections, the heavy mesons cannot be accelerated to acquire
such high velocities in the hadronic phase. 

The transverse flow will also affect the spectra of photons  and
dileptons\cite{kampfer}. We find \cite{future} that the yield of single 
photons with large transverse momenta increases
by a factor of almost 5 due to the transverse expansion, as compared to
the situation when there is only a longitudinal expansion. Similar
trends are also seen in the transverse mass distribution of dileptons.
We have seen that gluon fugacities are larger than the quark fugacities.
This provides that the annihilation processes 
($\propto \lambda_q \lambda_{\bar q}$), 
give only a small contribution as compared to the Compton processes
($\propto \lambda_g \lambda_q$) to the production of single photons, and
can be neglected. Dileptons on the other hand  are produced from 
quark-antiquark annihilation.  A comparison of these two measurements 
could be a valuable source of information on the quark and gluon 
fugacities of the system\cite{strickland}.

An experimental measurement of the strong transverse flow predicted here
would provide two important pieces of evidence about the properties of
the dense matter produced in ultrarelativistic heavy ion reactions:
that the matter is formed at very high initial pressure, and that this
pressure is maintained for a substantial period of time.
If the ``rarified'' partonic matter hadronizes before it attains
chemical equilibrium, as is predicted at RHIC energies, it is possible 
that a mixed phase or a interacting hadronic phase is not created.
The emitted hadrons would then reflect the thermal characteristics
of the quark phase from which they emerge.  We are currently investigating
whether another initial parton distribution would lead to a
significantly different transverse flow pattern that could be 
experimentally discriminated from the one predicted here.

{\it Acknowledgments:} We acknowledge useful comments from Dr Bikash Sinha.
This work was supported in part by a grant from the U.S. Department of
Energy (DE-FG02-96ER40945).

\begin{table}
\begin{center}
\begin{tabular}{|l|r|r|} 
$\tau_i=0.25$ fm/$c$ &RHIC &LHC \\ \hline
$\epsilon_i$ (GeV/fm$^3$) &61.4 &425 \\
$T_i$ (GeV) &0.668 &1.02 \\
$\lambda_g^{(i)}$ &0.34 &0.43 \\ 
$\lambda_q^{(i)}$ &0.064 &0.082 \\ 
\end{tabular}
\end{center}
\caption{Initial conditions for the hydrodynamical expansion phase
at RHIC and LHC.  The initial time is taken as $\tau_i=0.25$ fm/$c$.}
\end{table}

\begin{figure}
\epsfxsize=3.25in
\epsfbox{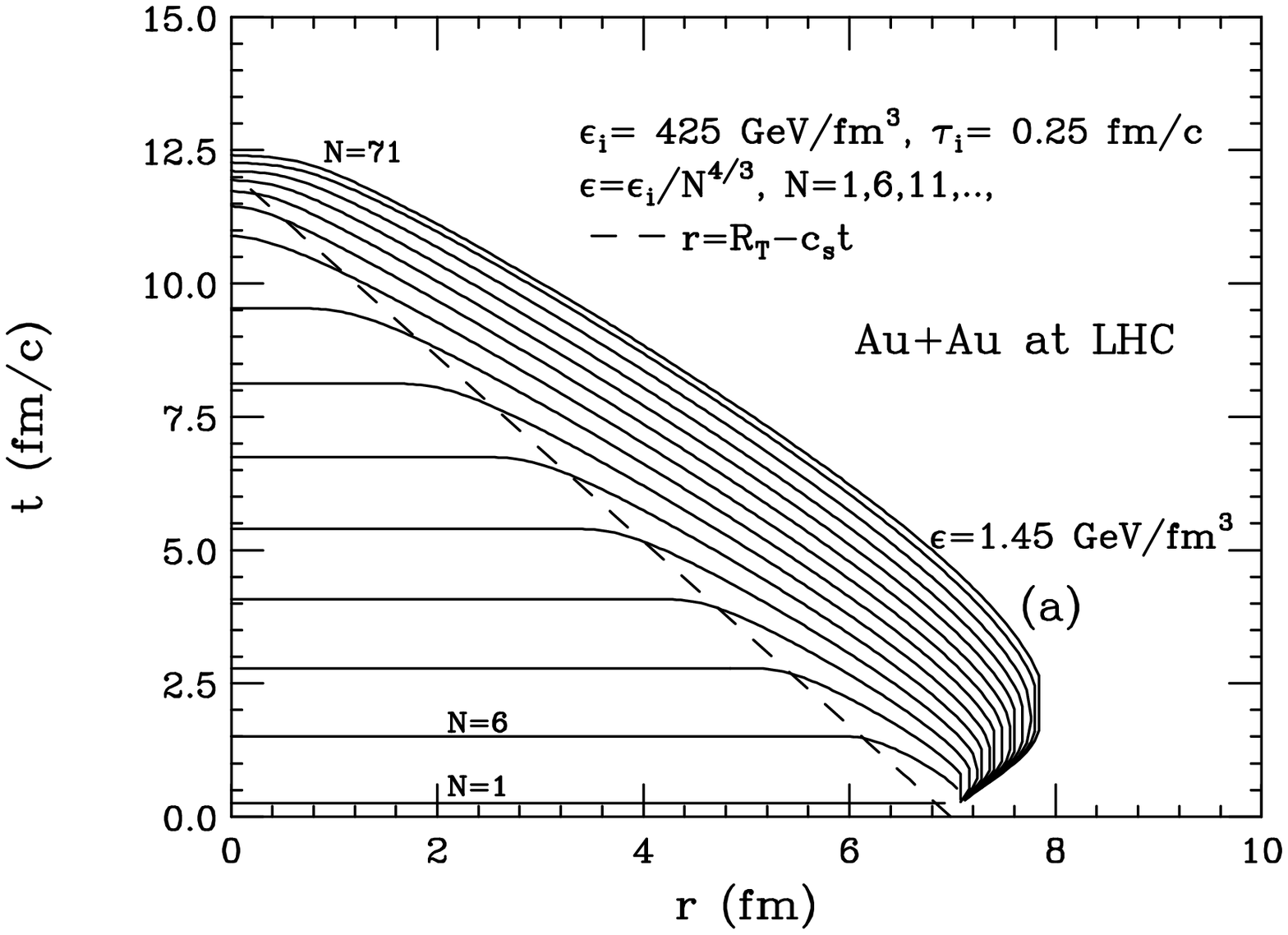}
\epsfxsize=3.25in
\epsfbox{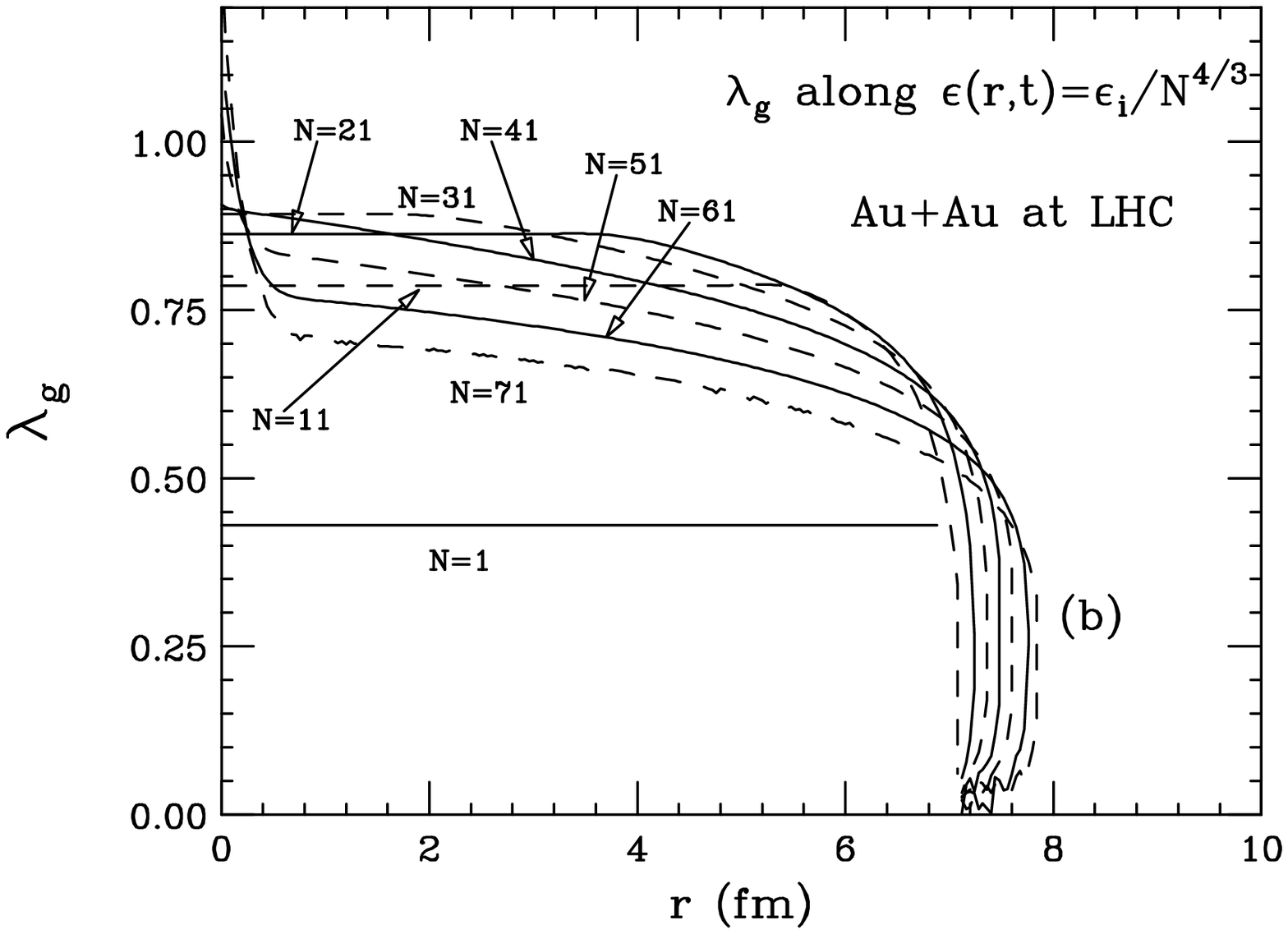}
\epsfxsize=3.25in
\epsfbox{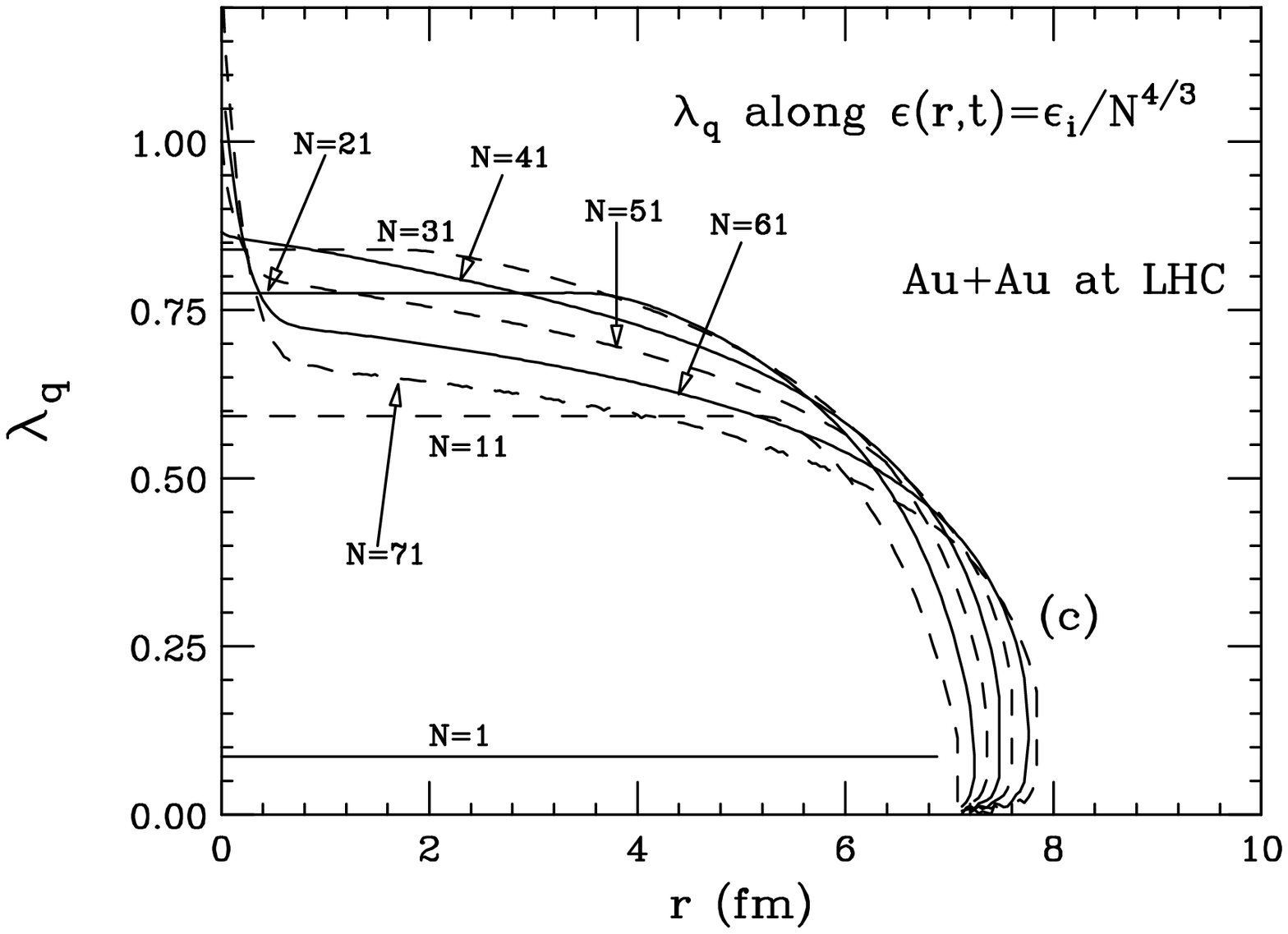}
\caption{(a) Constant energy density contours for a transversely
expanding quark-gluon plasma likely to be created in Au+Au collisions at
LHC. For a boost invariant longitudinal expansion these contours would be
equidistant ($\Delta t=5\tau_i$ at $r=$0) lines parallel to $N=1$. The 
dashed lines gives the distance from the axis where the rarefaction
wave has arrived at time $t$, and beyond which the fluid is strongly 
affected by the flow.
(b) Gluon fugacity along some of the constant energy density contours.
(c) Quark fugacities along some of the constant energy density contours.}
\end{figure}

\begin{figure}
\epsfxsize=3.25in
\epsfbox{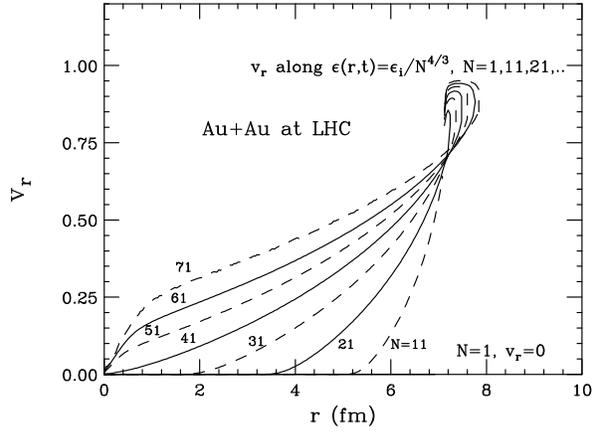}
\caption{Transverse velocities along some of the constant energy
density contours.}
\end{figure}

\end{document}